 \documentclass[final,5p,times,twocolumn]{elsarticle}

 \usepackage{graphics}

\usepackage{amssymb}

\journal{NIM A  RICAP-2013}

\begin{document}

\begin{frontmatter}



\title{ Perspectives of dark matter searches with antideuterons  }


\author[a,b]{A. Vittino}
\ead{vittino@to.infn.it}
\author[a,b]{N. Fornengo}
\ead{fornengo@to.infn.it}
\author[c,d]{L. Maccione}
\ead{luca.maccione@lmu.de}

\address[a]{Department of Physics, University of Torino \\ via P. Giuria 1, 10125 Torino, Italy}
\address[b]{Istituto Nazionale di Fisica Nucleare \\ via P. Giuria 1, 10125 Torino, Italy}
\address[c]{Ludwig-Maximilians-Universit\"{a}t, Theresienstra{\ss}e 37, D-80333 M\"{u}nchen, Germany}
\address[d]{Max-Planck-Institut f\"{u}r Physik (Werner Heisenberg Institut), F\"{o}hringer Ring 6, D-80805 M\"{u}nchen, Germany}

\begin{abstract}
The search for an excess of antideuterons in the cosmic rays flux has been proposed as a very promising channel for dark matter indirect detection, especially for WIMPs with a low or intermediate mass. With the development of the AMS experiment and the proposal of a future dedicated experiment, i.e. the General Antiparticle Spectrometer (GAPS), there are exciting possibilities for a dark matter detection in the near future.   
We give an overview on the principal issues related both to the antideuterons production in dark matter annihilation 
reactions and to their propagation through the interstellar medium and the heliosphere, with a particular focus on the impact of various solar modulation models on the flux at Earth. Lastly, we provide an updated calculation of the reaching capabilities for current and future experiments compatible with the constraints on the dark matter annihilation cross section imposed by the antiproton measurements of PAMELA.

\end{abstract}


\end{frontmatter}


\section{Introduction}
\label{sec:introduction}
Antideuterons have been proposed as a very promising channel for dark matter (DM) indirect detection in \cite{DFS}. Since DM particles annihilate (or decay) at rest, they are expected to produce a flux of antideuterons which, in the low energy region (i.e. below a few GeV/n), can be significantly larger than the astrophysical background (produced by the spallation of cosmic rays particles on the interstellar medium).  

The predicted $\bar{d}$ fluxes both for the DM signal and the secondary component are well below the current most constraining experimental upper limit given by the BESS collaboration \cite{Fuke:2005it}: $\phi_{\bar{d}}<1.9\times10^{-4}\:\mathrm{(m^2\: s\: sr\: GeV/n)^{-1}}$ in the energy interval (0.17-1.15) GeV/n at 95\% of confidence level. 
In the next years, this limit will be sensibly lowered since two experimental collaborations will look for antideuterons with a significantly improved sensitivity: the Alpha Magnetic Spectrometer (AMS) \cite{Battiston:2008zza,Incagli:2010zz,Bertucci:2011zz,Kounine:2012zz} is already under operation onboard the International Space Station since 2011, while the General Antiparticle Spectrometer (GAPS) \cite{Mori:2001dv,Hailey:2009zz,Hailey:2013gwa,Mognet:2013dxa,Fuke:2013lca} is a proposed experiment which is expected to begin its science flights in 2017/2018 from Antarctica. AMS will detect a minimal flux of $4.5\:\times\:10^{-7}\:\mathrm{(m^2\: s\: sr\: GeV/n)^{-1}}$ (corresponding to 1 detected event in a data taking period of 3 years) in the energy range (0.2-0.8) GeV/n \cite{giovacchini,choutko} while GAPS will be sensitive to fluxes as small as  $2.8\:\times\:10^{-7}\:\mathrm{(m^2\: s\: sr\: GeV/n)^{-1}}$ (corresponding to 1 detected events in the GAPS LDB+ which corresponds to a data taking period of 210 days) in the energy range (0.1-0.25) GeV/n \cite{haileypriv}.

In light of these exciting prospects on the experimental side, and of some recent theoretical developments (see, for example, \cite{strumia,ibarra,Dal:2012my}), we perform here a complete re-analysis of the predictions for the DM antideuterons flux (for a more complete analysis, we address the reader to Ref. \cite{Fornengo:2013osa}): in Section \ref{sec:production} we study the problem of $\bar{d}$ formation, while in Section \ref{sec:propagation} we explore the issues related to the propagation of antideuterons in the galactic and solar environment, with a particular focus on the solar modulation modeling. Lastly, in Section \ref{sec:prospects} we determine the prospects for a DM detection in the $\bar{d}$ channel and in Section \ref{sec:conclusions} we present our conclusions.  

\section{Antideuterons production}
\label{sec:production}
\begin{figure*}[t] 
\includegraphics[width=0.33\textwidth,height=0.23\textheight]{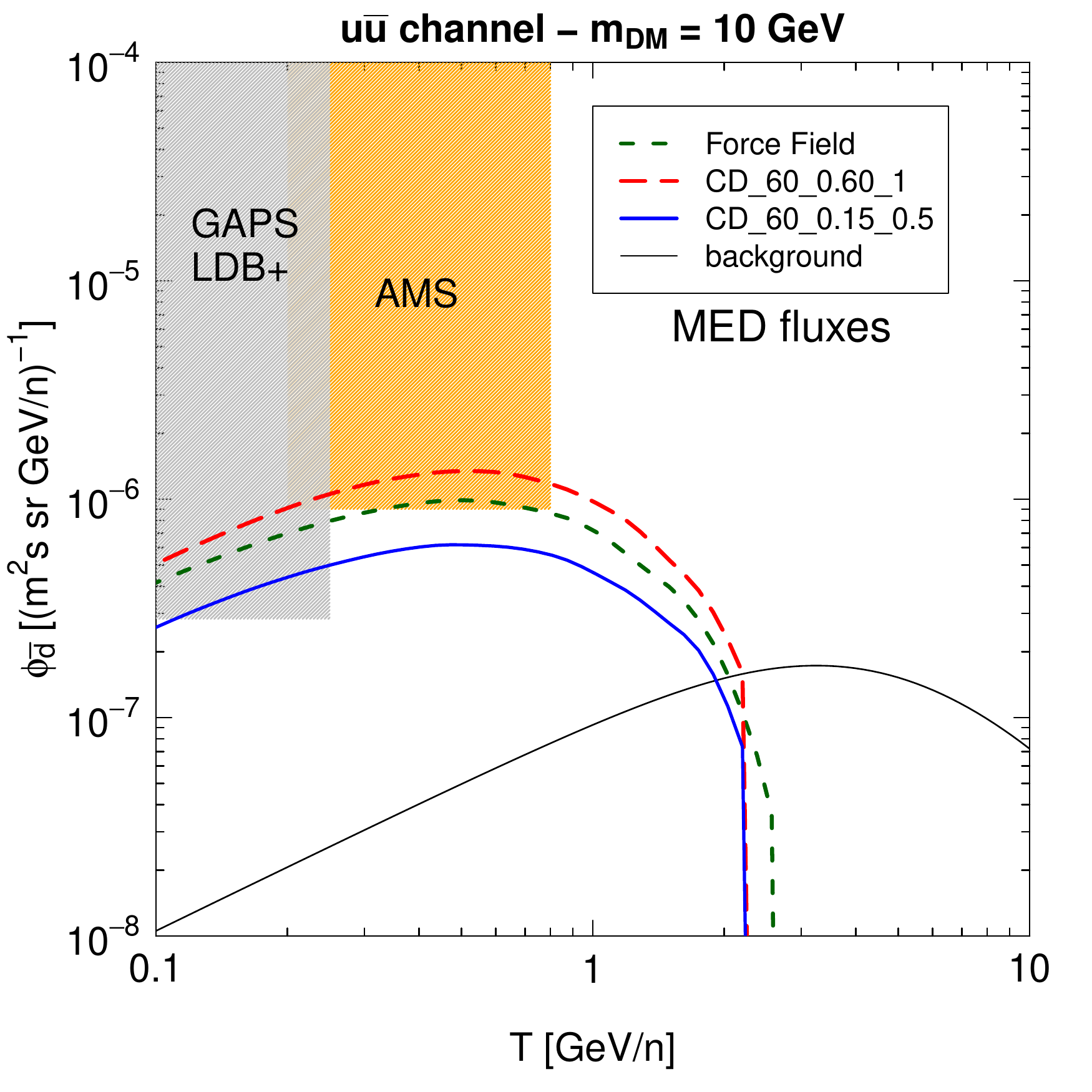}
\includegraphics[width=0.33\textwidth,height=0.23\textheight]{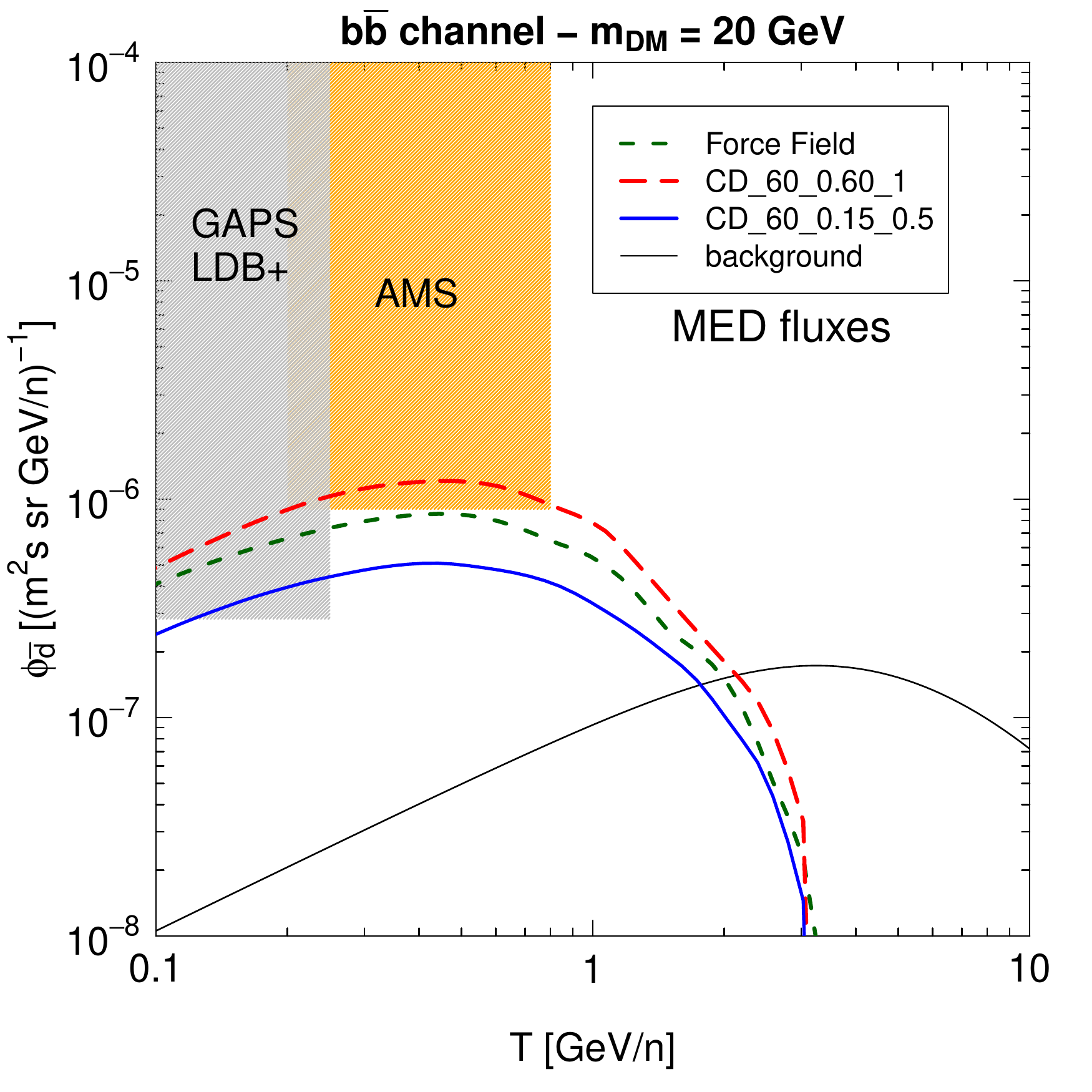}
\includegraphics[width=0.33\textwidth,height=0.23\textheight]{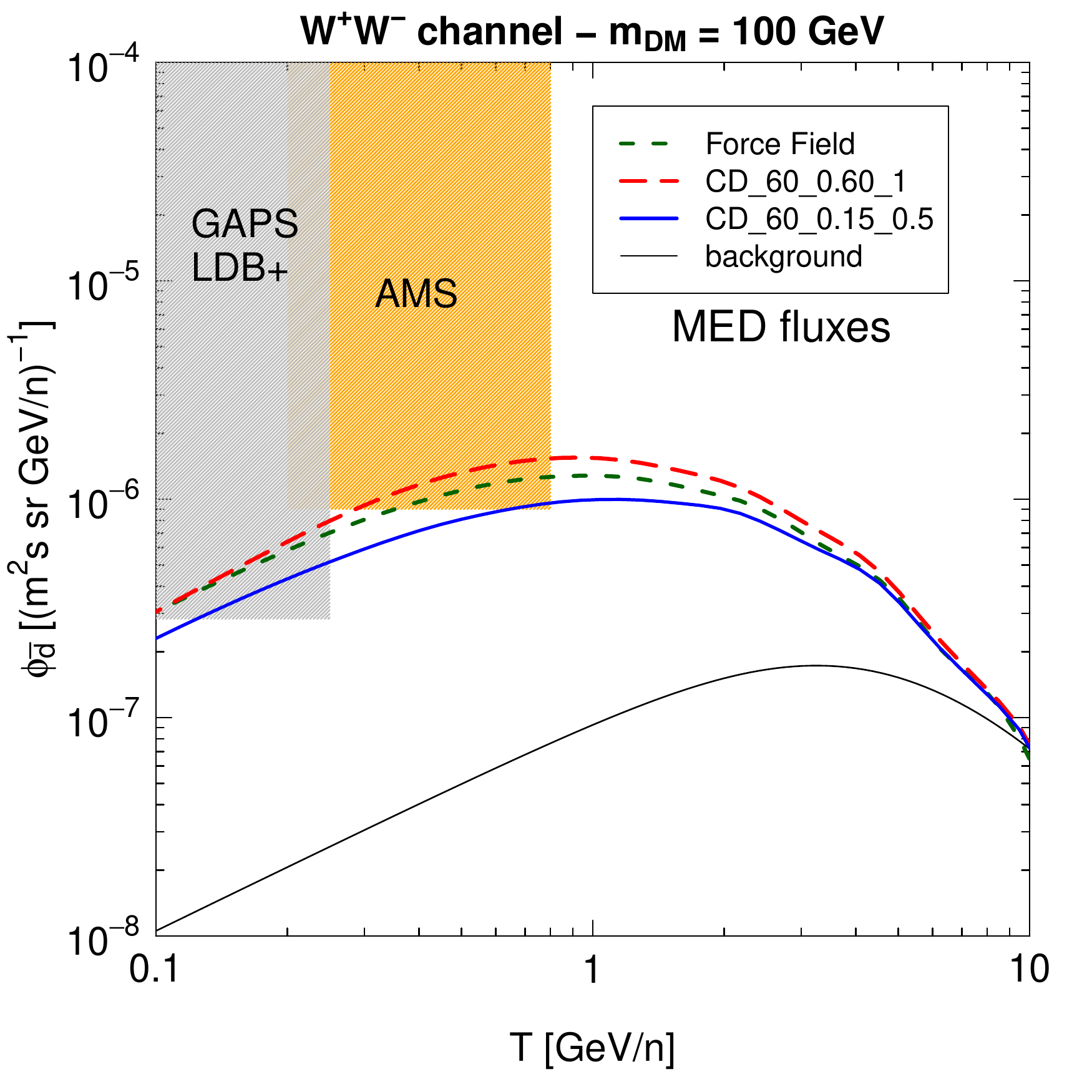}
\caption{Top-of-the-atmosphere $\bar{d}$ flux as a function of the $\bar{d}$ kinetic energy for 
three representative DM candidates: a 10 GeV DM annihilating into the $u\bar{u}$ channel 
with a cross section $<\sigma v>\:=\:2 \times 10^{-27}~{\rm cm^3s^{-1}}$ , a 20 GeV WIMP 
annihilating into the $b\bar{b}$ channel with a cross section $<\sigma v>\:=\:1 \times 10^{-26}~{\rm 
cm^3s^{-1}}$ and a DM particle annihilating into $W^+W^-$ with a cross section $<\sigma v>\:=\:6 
\times 10^{-26}~{\rm cm^3s^{-1}}$.  Several solar modulation models are considered (see \cite{Fornengo:2013osa} for an explanation of the codes used in the boxed insets). The two shaded regions 
correspond to the $3\sigma$ c.l. expected sensitivities for the experiments GAPS LDB+ and 
AMS-02, while the thin solid line denotes the astrophysical background (which is taken from \cite{DFM}). }
\label{fig:TOA}
\end{figure*}
Antideuterons are produced in a mechanism called coalescence \cite{coalescence1}. The idea behind coalescence is simple: two antinucleons produced in the same event can merge if they happen to be close enough in their phase space. If we define $F_{\bar{d}}$ as the phase space distribution of antideuterons, we can write:
\begin{equation}
 F_{\bar{d}} = \int  F_{(\bar{p}\bar{n})} (\sqrt{s},\vec{k}_{\bar{n}},\vec{k}_{\bar{p}}
)\; {\cal C}(\sqrt{s},\vec{k}_{\bar{n}},\vec{k}_{\bar{p}}
)\;
 d^3\vec{k}_{\bar{n}} \; d^3\vec{k}_{\bar{n}}
\label{eq:coa2}
\end{equation} 
Where $F_{(\bar{p}\bar{n})}$ is the momentum distribution of the $(\bar{p},\bar{n})$ pair and ${\cal C}(\sqrt{s},\vec{k}_{\bar{n}},\vec{k}_{\bar{p}})$ is the coalescence function, which represents the probability that the two antinucleons merge. In Refs. \cite{strumia,ibarra} it is shown that the presence of (anti)correlations between the two antinucleons can largely affect the $\bar{d}$ flux produced by DM annihilation. 
In order to take into account these correlations we use an event-per-event coalescence model to sample the $F_{(\bar{p}\bar{n})}$ distribution: we simulate the DM pair annihilation with a MC event generator (i.e. PYTHIA 6.4.26 \cite{Pythia}) and we calculate the relative momentum and physical distance of all the $(\bar{p},\bar{n})$ pairs present in the final state of each event;  if these two quantities are smaller than the two cut-off values $p_0$ and $R_*$ (we assume $R_*$  to be 2 fm which is the radius of the antideuteron; larger values for this parameter could in principle be assumed but, unless they are made unreasonably large, this would have only a marginal effect in the final results) we consider them as an antideuteron. 

In order to tune the coalescence momentum $p_0$, we compare the results predicted by our model with the experimental measurements related to the process that can be considered to possess more similarities with a DM pair annihilation, i.e. the inclusive reaction $e^+~e^-\rightarrow \bar{d}X$: for this process, the ALEPH collaboration at LEP has measured the $\bar{d}$ production rate at the Z resonance for antideuterons with a momentum in the range (0.62,1.03) GeV and a polar angle that satisfies the condition $|cos \theta| < 0.95$ \cite{Aleph}. The value of the coalescence momentum $p_0$ that reproduces this result in the framework of our coalescence model is: $p_0 = (195 \pm 22)\: \mathrm{MeV}$.

We must inform the reader, however, that this determination of the coalescence momentum $p_0$ is based on the comparison with only one experimental point and this does not make us able to investigate a possible dependence of the coalescence parameter with the center of mass energy of the process. This can, in principle, affect both the size and the spectral features of the DM signals that we calculate.

\section {Propagation in the galactic and solar environment}
\label{sec:propagation}

\begin{figure*}
\includegraphics[width=0.33\textwidth,height=0.23\textheight]{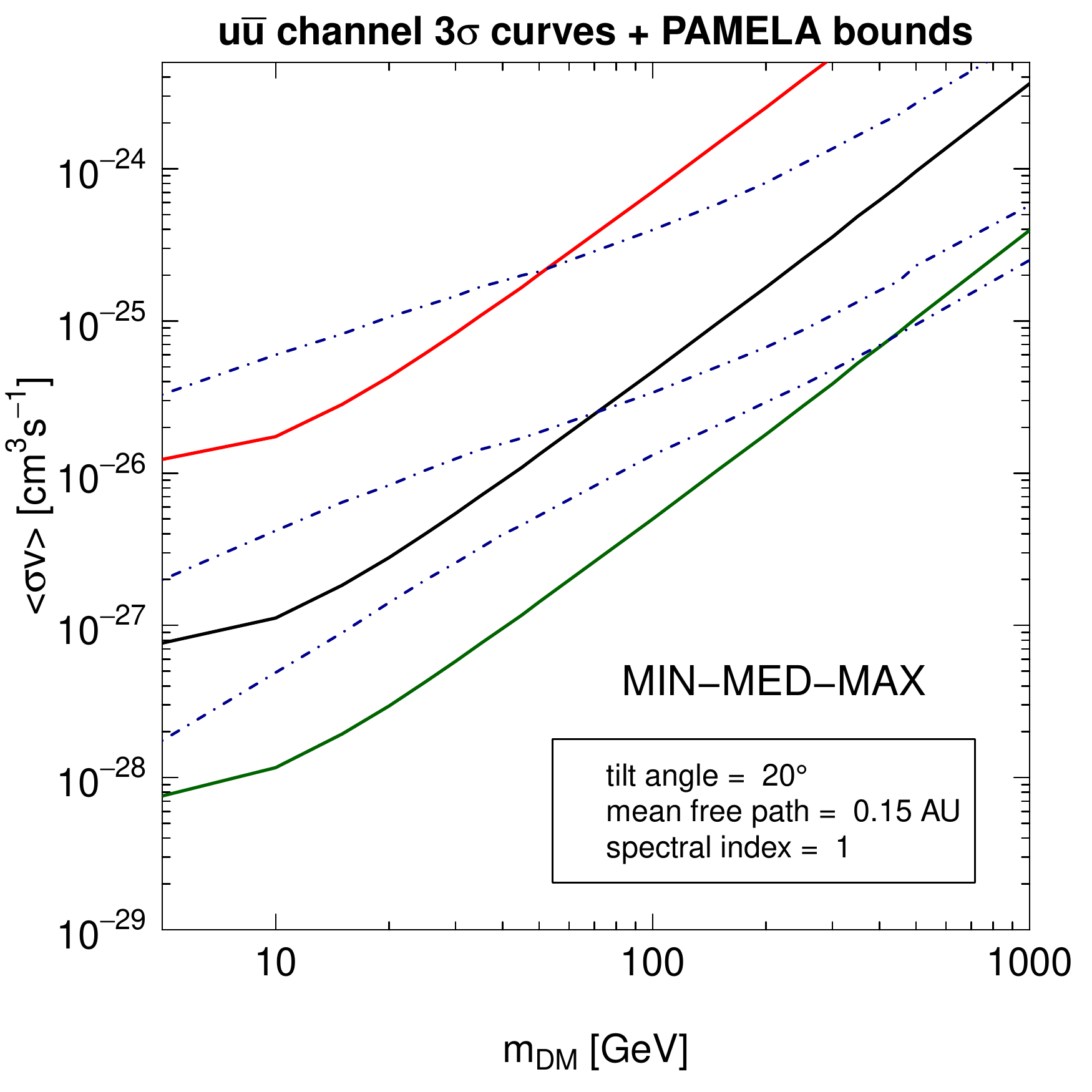}
\includegraphics[width=0.33\textwidth,height=0.23\textheight]{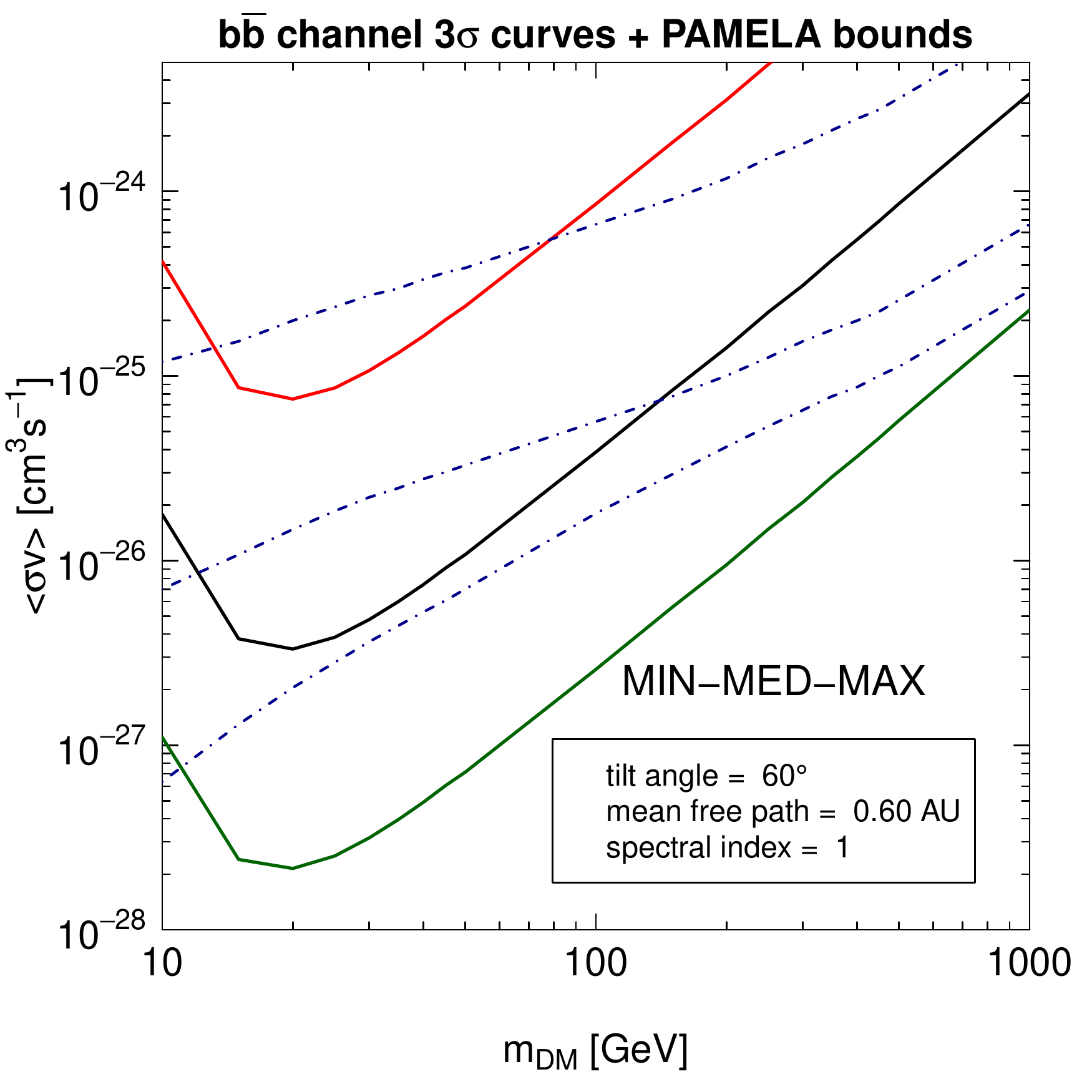}
\includegraphics[width=0.33\textwidth,height=0.23\textheight]{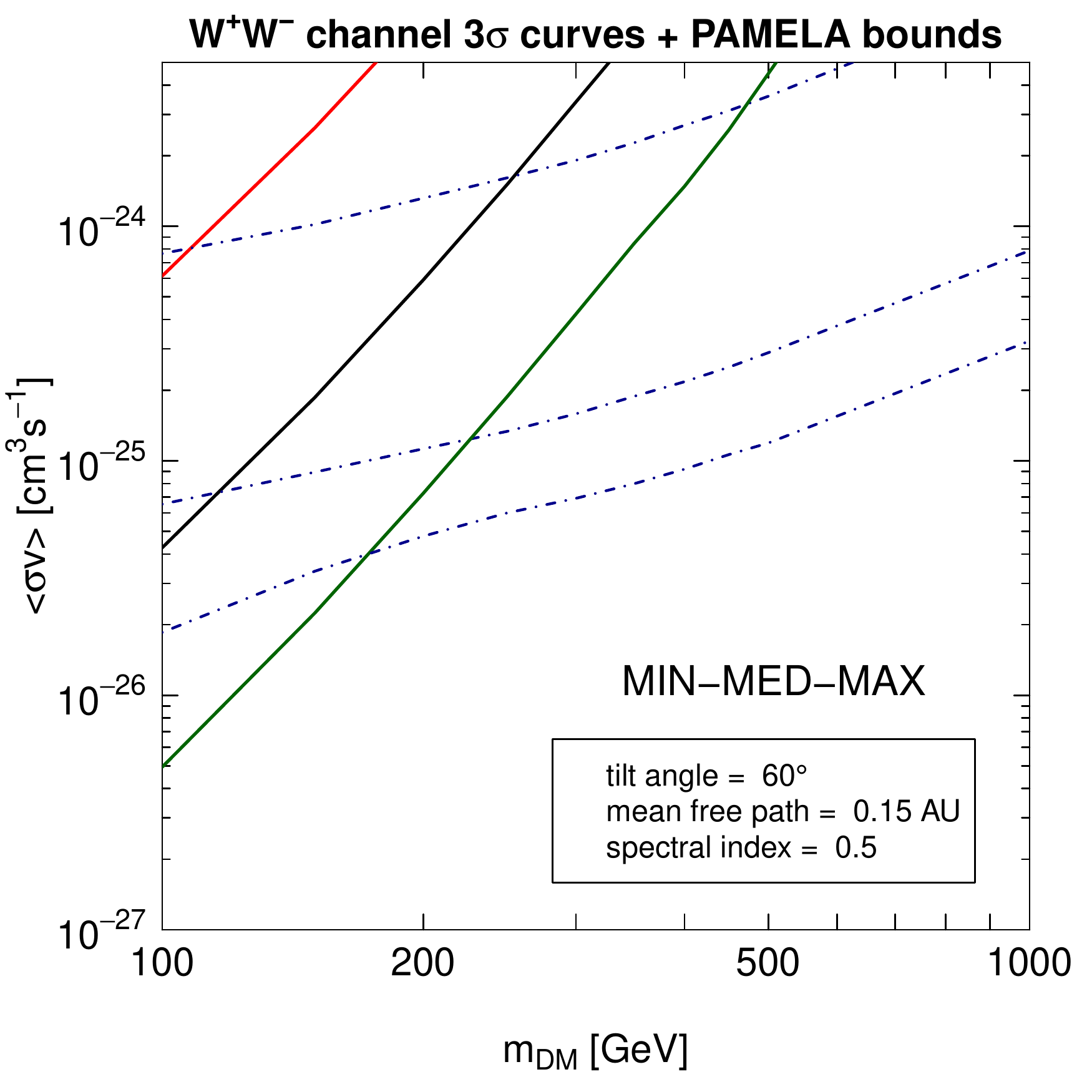}
\caption{Prospects for a 3$\sigma$ detection of a DM signal in the $\bar{d}$ channel for the GAPS experiment. The three upper/median/lower solid lines denote the reachability curves (which corresponds to the observation of 1 $\bar{d}$ event) for the three propagation models MIN, MED and MAX. Dot dashed lines show the corresponding bounds from PAMELA. The solar modulation models used are reported in the boxed insets.}
\label{fig:reach}
\end{figure*}
Once that they are produced, antideuterons propagate through the interstellar medium (ISM) and their propagation is usually described by a transport equation: 
\begin{equation}
-\nabla [K \nabla n_{\bar{d}}] + V_{c}\frac{\partial}{\partial{z}}n_{\bar{d}}+2h~ \delta (z)~\Gamma_{\rm ann}^{\bar{d}}n_{\bar{d}} = q_{\bar{d}}
\label{eq:transport}
\end{equation}
where $n_{\bar{d}}$ is the antideuterons number density, $K$ is the diffusion coefficient, $V_c$ is the velocity of the convective wind, $\Gamma_{\rm ann}^{\bar{d}}$ is the $\bar{d}$ interaction rate with the hydrogen and helium nuclei that populate the ISM and $q_{\bar{d}}$ is the $\bar{d}$ source term which, for the pair annihilation of a DM particle with mass $m_{DM}$, can be written as:
\begin{equation}
 q_{\bar{d}}(r,z,E) = \frac{1}{2} <\sigma v> \frac {dN_{\bar{d}}}{dE}\left(\frac{\rho(r,z)}{m_{DM}}\right)^2
\end{equation} 

Where $<\sigma v>$ is the thermally averaged DM annihilation cross section, $ dN_{\bar{d}}/dE$ is the $\bar{d}$ injected spectrum and $\rho(r,z)$ is the DM density profile. In this work we will always assume an Einasto profile: 
\begin{equation}
\rho(r,z)/\rho_\odot = \exp(-2[(\sqrt{r^2+z^2}/r_s)^{\alpha}-(r_{\odot}/r_s)^{\alpha}]/\alpha)
\end{equation}
with $\alpha = 0.17$ and $r_s$ = 20~kpc.

We solve Eq. \ref{eq:transport} in the simplified framework of the two-zone diffusion model \cite{diffusion1,diffusion2,diffusion3}, which is based on the two assumptions that the diffusion is confined inside a cylinder of radius $R ~= ~20~ \mathrm{Kpc}$ and half-thickness L out of the galactic plane and the interaction with the interstellar medium can only take place in a disk of vertical half-height $h\:=\:100~{\rm pc}$ coincident with the galactic plane. In addition, we also assume a diffusion coefficient which is only energy dependent: $K(r,z,E) = \beta K_0 \left( {{\cal R}} /{1~\mbox{GV}}\right)^\delta$ and a constant convective velocity $V_c$. Therefore, our propagation model is identified by the values of the parameters (L, $K_0$, $\delta$, $V_c$) which are usually obtained through the study of B/C data. We adopt here the three reference sets of parameters usually called MIN, MED and MAX \cite{minmedmax} which are the ones typically used in literature.

When they enter the heliosphere, cosmic rays (CRs) diffuse through the solar magnetic field (SMF) which has a polarity that changes every 11 years and it has the form of a Parker spiral:
\begin{equation}
\vec{B} = AB_{0}\left(\frac{r}{r_{0}}\right)^{-2}\left(\hat{r} - \frac{\Omega r\sin\theta}{V_{\rm SW}}\hat{\varphi}\right)\;,
\end{equation} 
being $\Omega$ the differential rotation rate of the Sun, $V_{\rm SW}$ the solar wind velocity, $\theta$ the colatitude and $B_0$ a normalization constant such that $|B|(1~{\rm AU})=5~{\rm nT}$. The function $A=\pm H(\theta-\theta')$ describes the SMF polarity and the Heaviside function $H$ is used to take into account the presence of the Heliospheric Current Sheet (HCS) which is a geometrical surface that separates magnetic field lines according to their polarity. The geometry of the HCS is fully described by the function $\theta' = \pi/2 + \sin^{-1}\left(\sin\alpha\sin(\varphi+\Omega r/V_{\rm SW})\right)$ being  $\alpha$ known as the tilt angle.  The modulation effects which affect CRs during their propagation in the heliosphere can be described by the following transport equation \cite{1965P&SS...13....9P}:
\begin{equation}
\frac{\partial f}{\partial t} = -(\vec{V}_{\rm sw}+\vec{v}_d)\cdot \nabla f + \nabla\cdot ({\bf K}\cdot\nabla f) + \frac{P}{3}(\nabla\cdot\vec{V}_{\rm sw})\frac{\partial f}{\partial P}\;,
\label{eq:solartransport}
\end{equation}
where $f$ is the CR phase space density (averaged over momentum directions), ${\bf K}$ is the symmetrized diffusion tensor, $\vec{v}_d$ is the divergence-free velocity associated to drifts and $P$ is the CR momentum. We assume the diffusion to be present only in the direction parallel to the magnetic field lines and for the parallel CR mean free path we take $\lambda_{\|} = \lambda_{0}(\rho/1~{\rm GeV})^{\gamma}(B/B_{\bigoplus})^{-1}$ being $B_{\bigoplus}\:=\:5~{\rm nT}$ the magnetic field at the Earth position \cite{2011ApJ...735...83S}. 

We exploit the recently developed code $\textsc{HelioProp}$ \cite{Maccione:2012cu} in order to explore various configurations of the solar modulation parameters (i.e. $\alpha$, $\lambda_0$ and $\gamma$).

\section{Prospects for DM detection}
\label{sec:prospects}

Since a DM annihilation event which produces antideuterons is also assumed to produce a much larger amount of antiprotons, the measurements of the antiprotons flux performed by the PAMELA experiment \cite{Adriani:2010rc} will strongly constrain the possible annihilation cross section of our DM candidate. We determine the antiproton bounds by performing a full spectral analysis of the whole set of PAMELA data (i.e. from 60 MeV to 180 GeV in terms of kinetic energy). Our bounds are calculated at a 3$\sigma$ confidence level and they are obtained by imposing a 40\% uncertainty to the astrophysical background (for which we use the one calculated in \cite{Donato:2008jk}). For the solar modulation used in the calculation of the bounds, we use a set of parameters compatible with the data taking period of PAMELA: $\alpha=20^\circ$, $\lambda_0 \: =\: 0.15~{\rm AU}$, $\gamma\:=\:1$.

\begin{figure*}
\centering
\includegraphics[width=0.33\textwidth]{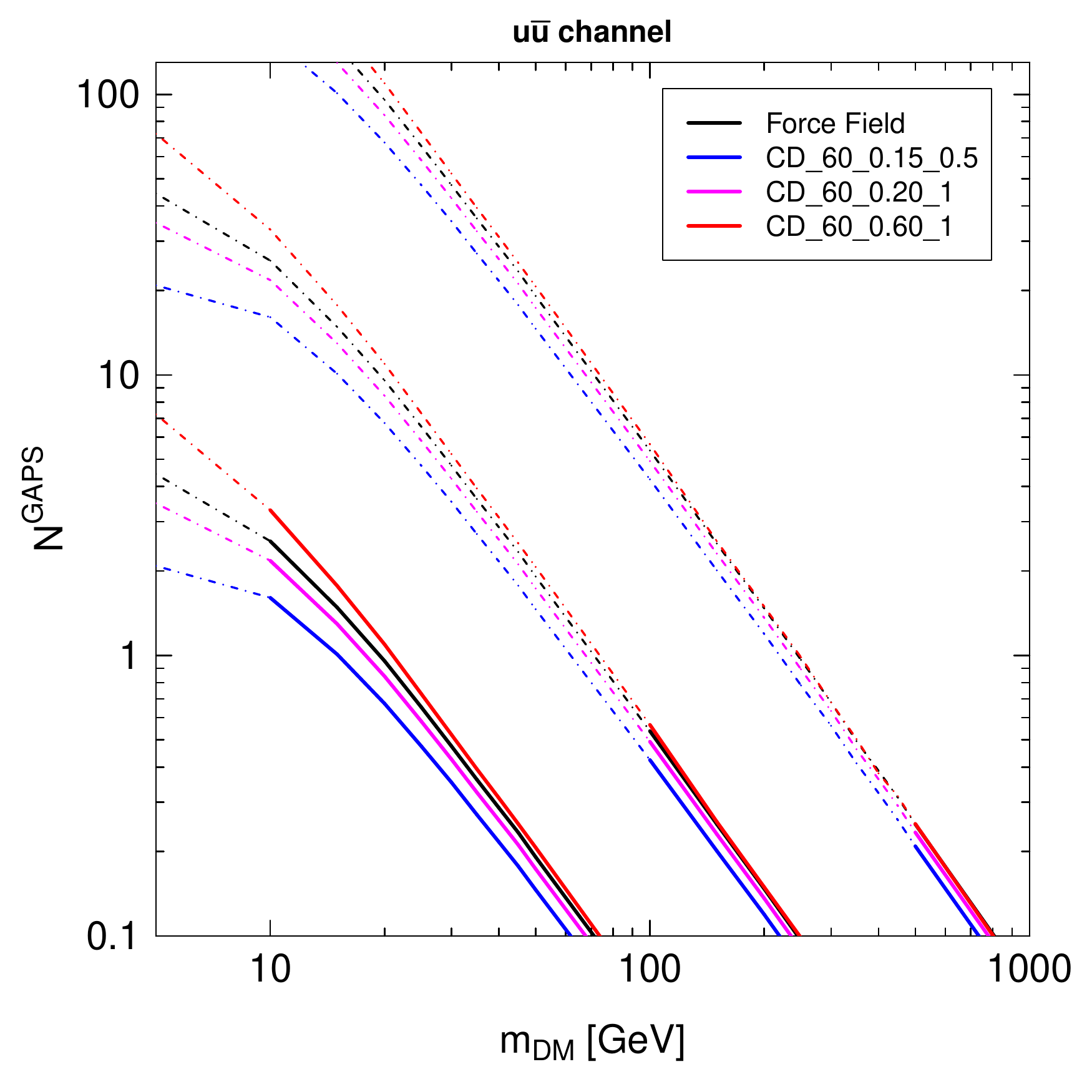}
\includegraphics[width=0.33\textwidth]{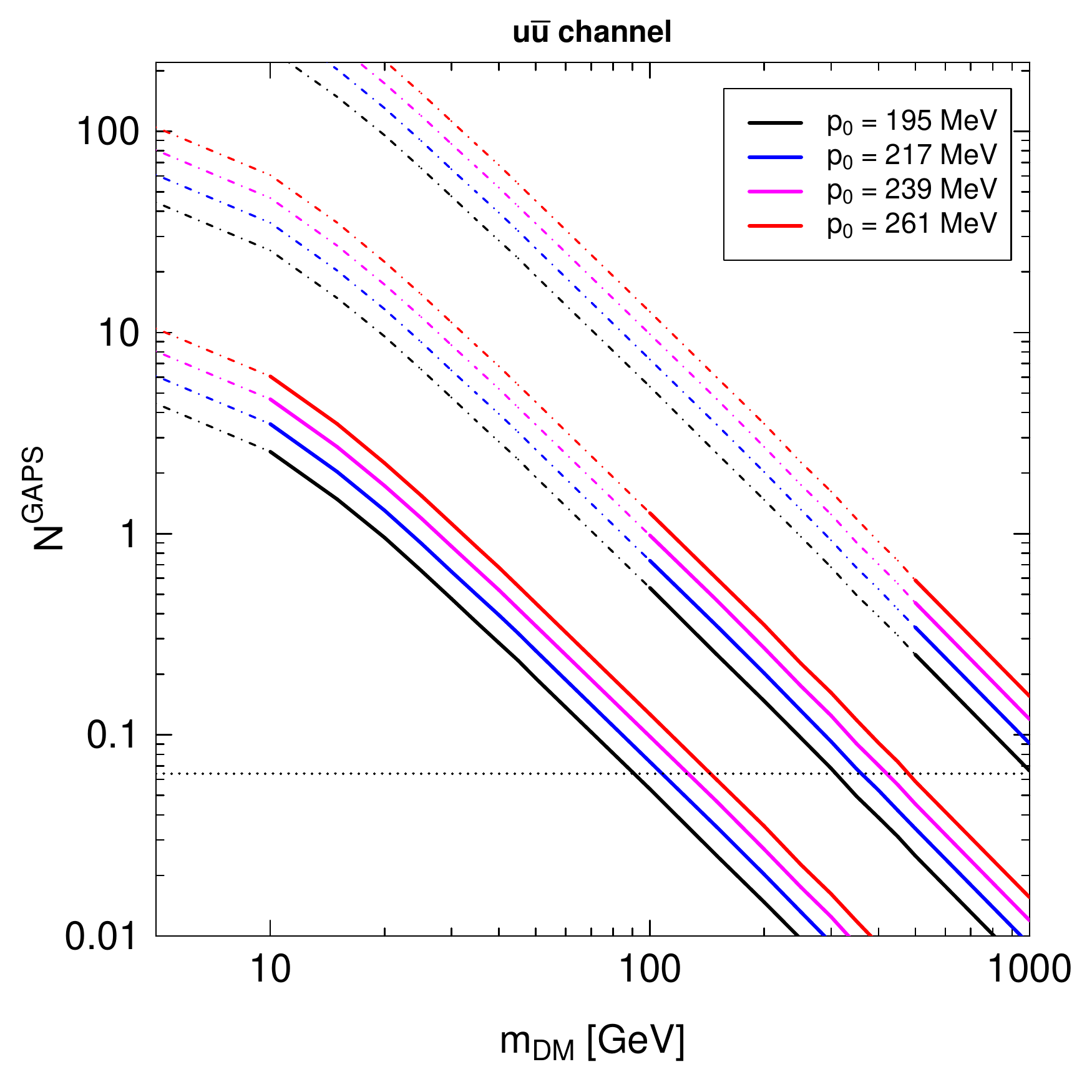}
\includegraphics[width=0.33\textwidth]{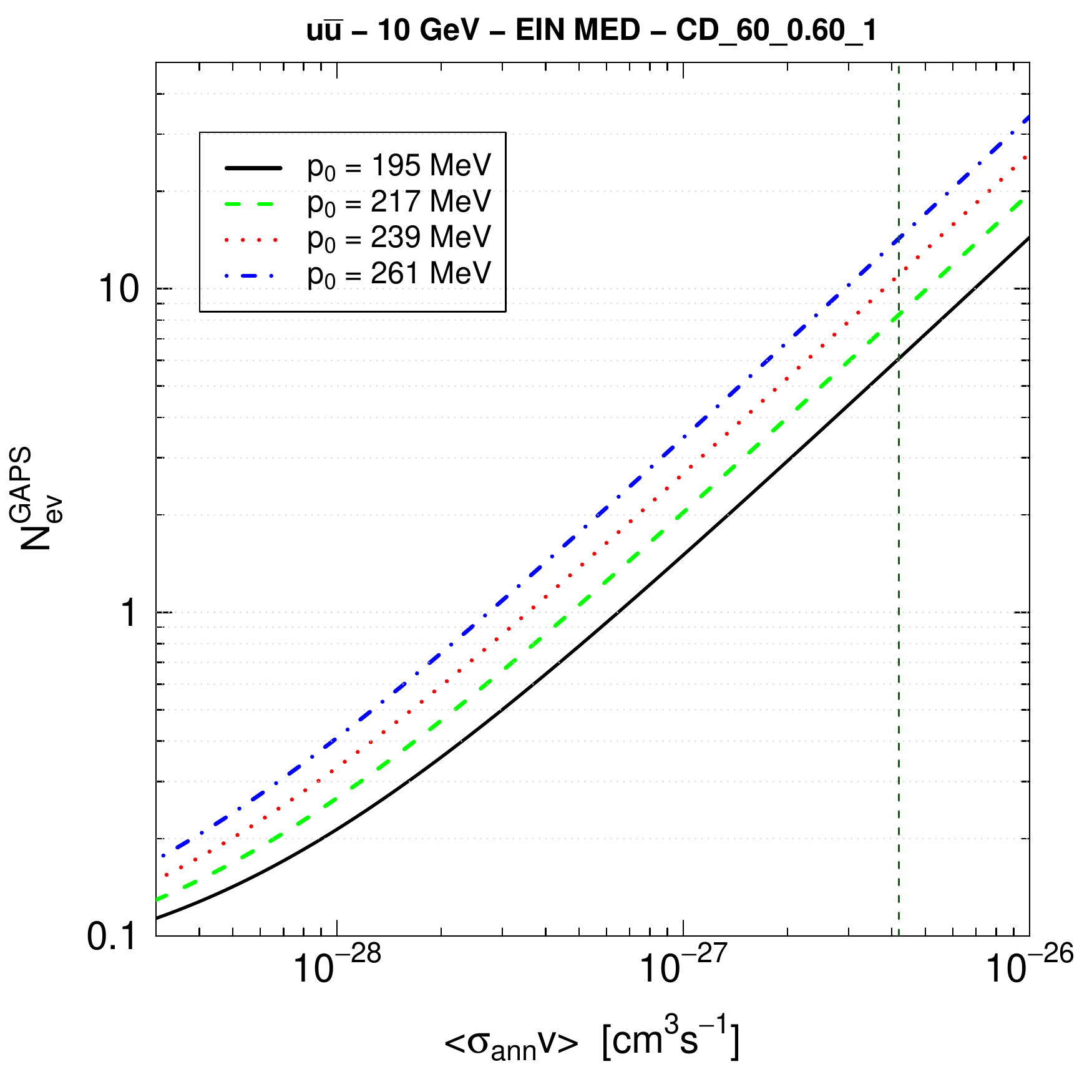}
\caption{Number of $\bar{d}$ events in the $u\bar{u}$ annihilation channel for the GAPS experiment. In the left and central panels, the three set of curves denote three different values of $<\sigma v>$: 0.1, 1 and 10 times the thermal value of $2.3\:\times\:10^{-26}~{\rm cm^3~s^{-1}}$. Solid (dot-dashed) lines are for configurations compatible (not compatible) with the PAMELA bounds. In the left panel various solar modulation models are considered, while in the central panel the modulation is done with a simple force field approximation and the coalescence parameter $p_0$ is varied inside its 3$\sigma$ allowed region. The horizontal black line represents the number of background events. In the right panel, the mass and the solar modulation are constant and different values of $p_0$ are considered}
\label{fig:number}
\end{figure*}

Even if the bounds that we derive are very constraining, we can see in Fig. \ref{fig:TOA} that our signals, obtained with annihilation cross sections compatible with PAMELA bounds, are expected to be largely in excess of the background flux and at the reach of both GAPS and AMS-02. In particular, perspectives for a detection appear to be particularly favourable for a light DM candidate annihilating in quarks pair (both $b\bar{b}$ and $u\bar{u}$), but even in the case of a 100 GeV DM particle that goes into a $W^+W^-$ pair, chances for a detection are still good. One can also easily see that solar modulation affects the final predicted flux at Earth, in particular in the low energy range, by a factor close to 2.   
In Fig. \ref{fig:reach} we show the reachability curves for the GAPS experiment in the $\left( m_{DM},<\sigma v> \right)$ plane, together with antiprotons bounds coming from PAMELA for the three annihilation channels $u\bar{u}$, $b\bar{b}$ and $W^+W^-$. We define the reachability curves as the set of configurations in the DM parameter space that make possible the observation of a number of antideuterons sufficient to claim for a DM detection with a $3 \sigma$ confidence level (this number, for the GAPS experiment is equal to 1). We can see that in the variety of galactic propagation and solar modulation frameworks considered, GAPS will be able to detect a DM signal in a large portion of the allowed parameter space. 
Lastly, in the three panels of Fig. \ref{fig:number}, we plot the number of expected events for the $u\bar{u}$ annihilation channel, for various annihilation cross sections, DM masses and coalescence momenta: we can notice that in the GAPS LDB+ mission, for this channel we can expect a number of signal events close to 15.  
\section{Conclusions}
\label{sec:conclusions}
We have seen that, despite the bounds imposed by antiprotons measurements, the detection of antideuterons is on the reach of current and future experiments for a wide variety of DM candidates and astrophysical configurations. We have shown that, even if the dominant role in the uncertainties affecting the calculated fluxes is played by galactic propagation, the modelization of the solar modulation seems to have a non negligible effect. Lastly, a detailed calculation of prospects for detection cannot ignore the role of the production mechanisms which are not fully understood yet.   





\bibliographystyle{elsarticle-num}
\bibliography{<your-bib-database>}

\end{document}